\documentclass[pra,aps,a4paper,twocolumn,10pt,floatfix]{revtex4-2}

\usepackage{amsmath,hyperref,graphicx,xcolor,amssymb}
\usepackage{amssymb}
\usepackage[caption=false]{subfig}
\usepackage{tikz}

\begin{document}

\title{Flow of non-classical correlations in cluster states due to projective measurements} 

\author{Chandan Mahto}
\email{chandanmahto00716@iisertvm.ac.in}
\affiliation{School of Physics, IISER Thiruvananthapuram, Kerala, India 695551}
\author{Anil Shaji}
\email{shaji@iisertvm.ac.in}
\affiliation{School of Physics, IISER Thiruvananthapuram, Kerala, India 695551}
\affiliation{Centre for High Performance Computing , IISER Thiruvananthapuram, Kerala, India 695551}

\begin{abstract}
We explore the flow of quantum correlations in cluster states defined on ladder type graphs as measurements are done on qubits located on the nodes of the cluster. We focus on three qubits at the end of the ladder and compute the non-classical correlations between two of the three qubits as measurements are done on the remaining qubits. We compute both the entanglement between the two qubits as well as the quantum discord between them after the measurements. We see that after all but three qubits are measured, the non-classical correlations developed between two of them show a trend of being stronger with the length of the ladder. It is also seen that measurements on to the basis states of operators belonging to the Clifford group do not produce such correlations or entanglement. The non-classical correlations produced depend only on the number, location and nature of preceding non-Clifford measurements. Our results not only throw light on the dynamics of quantum correlations while an algorithm proceeds step-by-step in the Measurement-based Based Quantum Computing (MBQC) model but it also reveals how the last two qubits, treated as an open quantum system, can have increasing entanglement or other non-classical correlations as its immediate environment is interrogated through random measurements.
\end{abstract}
\maketitle

\section{Introduction}

Quantum information, unlike its classical counterpart, can lie delocalized across multiple physical systems. Quantum entanglement and more generic non-classical correlations that are quantified by measures like quantum discord are the underlying physical resources that allow quantum information to take on delocalized forms. The ability to store and manipulate information delocalized across a qubit register is a key difference between quantum and classical information processing. The significance of this difference becomes evident when the resources that enable quantum computers to perform certain tasks exponentially faster than the best-known classical approach are traced out. For instance, among the various quantum resources that have been identified as being useful for quantum information processing~\cite{Chitambar:2019ko, Quantum_speedup, Coh_mix_resource, Coh_ent_resource, Stabilizer_resource, Coherence_resource, Quantum_resource}, it is known that for pure state quantum computing, multipartite entanglement that grows with the problem size is an essential requirement for exponential speedup~\cite{Jozsa2011}.  A comprehensive picture of the resources that facilitate a quantum advantage in information processing tasks beyond the pure state scenario is still not available but considerable progress has been made in this direction~\cite{bacon2009too, gour2024resources, salazar2024resource, berk2021resource, lostaglio2019resource, bhattacharya2020resource}.

Following the dynamics of non-classical correlations while a quantum computation progresses is the focus of this Paper. Such dynamics, particularly that of quantum entanglement, has been considered in various contexts previously~\cite{zyczkowski_dynamics_2001, ptaszynski_system-bath_2024, dahbi_dynamics_2022, saghafi_focusing_2019, aolita_open-system_2015}. Here we look at it using cluster states and the measurement based quantum computation (MBQC) model. The MBQC model deviates from the conventional circuit model by initializing computations with cluster states as entangled resource states arranged on specific lattices or graphs, followed by the execution of local measurements~\cite{Raussendorf:2001js, raussendorf_measurement-based_2003, briegel_measurement-based_2009}. Local unitary operations on a single qubit cannot affect information delocalized across the multiple qubits that form an entangled cluster state. However, generic quantum operations that are not unitary applied locally on a single qubit can modify the non-classical correlations it shares with other qubits in the cluster. Measurements are a special class of such operations which are of interest to us in the following since it detaches the measured qubit from the rest of the cluster while at the same time modifying the delocalized information shared across the remaining qubits. We look at the dynamics of specific non-classical correlations that are shared between a few qubits of a cluster state as measurements are performed on the remaining qubits. 

The cluster state, which is the starting point of MBQC, is also a stabilizer state. From the Gottesman-Knill theorem it is known that stabilizer states, while being highly entangled, can also be simulated efficiently using classical means and so by themselves cannot yield quantum computational advantage~\cite{gottesman1997stabilizer,aaronson_improved_2004}. Measurements on the individual qubits of the cluster not only remove the respective qubits from the cluster, they also can move the state of the remaining qubits away from being stabilizer states. These measurements modify the quantum information shared across all the qubits in a manner that allows for implementation of quantum algorithms that are exponentially faster than the best known classical ones on the cluster. Exploring how the delocalized information in cluster states respond to the measurements done on each of the qubits of the cluster is the subject matter of this Paper. We investigate this dynamics for simple, ``ladder"-shaped cluster states and focus on the non-classical correlations that develop between two qubits in the last rung of the ladder as measurements are done on other qubits in the cluster. 

In~\cite{SubsystemDiscord} we investigated the relationship between non-classical correlations in a subsystem of a globally entangled state and genuine multipartite entanglement in the global state. We found that when the global state of a collection of qubits considered has a regular structure like a cluster state, the two are typically related. Curiously, precisely when the the global state is a stabilizer state, the subsystem discord vanishes even if the global state is highly entangled. We take this line of investigation forward here, but in a slightly different direction. The ladder state is initially a stabilizer and the two qubits in the last rung share no quantum discord among them. We look at the non-classical correlations that develop between these two qubits as all the qubits in the cluster except the last three are measured. The measurements on the remaining qubits are chosen such that after each measurement, the state of the remaining qubits is pushed further away from being a stabilizer state. We choose to focus on the last three qubits of the ladder and study the quantum discord and entanglement between two of them because if the third one is also measured, the state of the remaining two will be pure and entanglement will be the only form of non-classical correlations present. With the third qubit also left unmeasured, there can be other forms of non-classical correlations between the last two. 

We can also view the last two qubits as an open quantum system coupled to a rather idealised environment that forms the rest of the ladder-like cluster state. From this perspective, we see how non-classical correlations can be built up in the open system due to measurements of a random nature happening on the environment that it is coupled to. As suggested in~\cite{SubsystemDiscord}, the presence of non-classical correlations in a mixed state quantum computer can then be viewed as a consequence of computationally useful (non-stabilizer) entanglement present in the pure state of a larger system into which the mixed state of the computer is embedded in. The simple ladder-state model allows us to explore quantitatively how such correlations change in response to random measurements occurring on the environment.  

This paper is organized as follows. In Sec.~\ref{clustersec} we review cluster states and their properties including non-classical correlations in subsystems. Flow of non-classical correlations in the ladder state model is discussed in the next section. Our conclusions are in Sec.~\ref{conclusion}.

\section{cluster States and MBQC \label{clustersec}}

A cluster state~\cite{PersistentofEntanglement} can be defined on a simple, un-directed graphs in any dimension. Qubits in the state $|+\rangle = (|0\rangle + |1\rangle)\sqrt{2}$ are placed on each of the vertices and each of the edges correspond to a two-qubit entangling gate $U_{kl}$ performed on the qubits placed on the vertices, $k$ and $l$ connected by the edge. The entangling gate used is typically the {\sc Controlled}-Z (CZ) gate. Cluster states are a special case of the more generic graph states~\cite{Hartmann:2007db, hein2006entanglement} wherein the initial states of the qubits on the vertices as well as the entangling operations done across the edges need not be uniform. 

A cluster state can be constructed from a collection of qubits all in the state $|0\rangle$ through the application of the Hadamard and CZ gates only. Since these operations form a subset of the Clifford group~\cite{calderbank1998CliffordGroup} of operations on a collection of qubits, the resultant cluster state is a stabilizer state~\cite{raussendorf_measurement-based_2003}. A stabilizer state $|\psi_{\rm stab}\rangle$ on $n$ qubits is the simultaneous eigenstate with $+1$ eigenvalue of $2^n$ operators, all of which are strings of Pauli operators (elements of the Clifford group). This allows for a succinct description of the state in terms of $n$ {\em generators} of the $2^n$ operators. Application of operators belonging to the Clifford group on such states results in states whose stabiliser generators can be computed directly from the generators of the initial state. In other words, Clifford group operations on stabilizer states produce other stabilizer states and any quantum information processing task restricted to stabilizer states and Clifford group operations can be efficiently simulated using classical means, as proven in the celebrated Gottesman-Knill theorem~\cite{gottesman1998heisenberg, aaronson_improved_2004}. 

\subsection{Measurement based quantum computing \label{mbqc}}

MBQC represents a distinctive approach to quantum computation that diverges from the traditional gate-based model. Gate-based quantum computing proceeds via application of quantum gates acting on a register of qubits, thereby evolving the state of the register towards a designated final state from which the output of the computation can be obtained~\cite{barenco1995elementaryGate, nielsen2002quantum}. MBQC, however, takes a different path that rely on measurements for implementing the quantum computational paradigm. Instead of applying quantum gates directly, MBQC utilizes a cluster state as the substrate for its operations~\cite{Nielsen}. Cluster states are an ideal starting point in that it provides a high degree of quantum entanglement to be used up as a resource as the computation proceeds. Given the graph corresponding to the desired cluster, the steps to be followed to prepare the cluster state is straightforward as well even if practical implementation can be challenging~\cite{kiesel2005experimentalCluster, zhang2006experimentalCluster}.  Entanglement in a cluster state is remarkably resilient against local operations, making it more challenging to disrupt than other highly entangled states like Greenberger-Horne-Zeilinger (GHZ) states. In fact, a single local operation cannot annihilate all the entanglement in a multipartite cluster state, while the same is not true for a GHZ state~\cite{PersistentofEntanglement,gisin1998GHZ}. 

MBQC is typically implemented on cluster states defined by simple and regular graphs like a two dimensional square lattice. Clusters defined on 2D triagonal, hexagonal and Kagome lattices are also known to be universal resource states for MBQC~\cite{VandenNest:2006he}. In fact, highly entangled states that do not necessarily share the symmetry and robustness of entanglement of the clusters defined on these simple lattices are known to be not useful for MBQC~\cite{TooEntangledGrossFlammia,TooEntangledBremner}. Starting from simple graphs like a 2D square lattice has the added advantage that given any quantum computation implemented through a gate-based quantum circuit, the same circuit can literally be ``drawn" on a suitably large cluster state showing that both models are equivalent. Drawing such a circuit involves removal of several qubits from the cluster which is achieved by measuring all of them on to the eigenbasis of the Pauli-$Z$ operator (computational basis). Using simple and symmetric graphs for building the cluster is also helpful for writing down the stabilizer generators for the cluster. The $n$ generators are identified by associating a string of Pauli operators to each of the $n$ qubits in the lattice which consists of the Pauli-$X$ operator on the qubit along with $Z$ operators on all the qubits connected to it.  

Once the superfluous qubits that are not part of the circuit drawn on the cluster are removed through measurements in the $Z$ basis, the computation proceeds through a sequence of measurements typically implemented from left to right on a square lattice. These local measurements are projections on to the eigenbasis of operators of the form,
\begin{equation}
  \label{measurementOP}
  O_{j}(\varphi_{j})=\cos{\varphi_{j}}X + \sin{\varphi_{j}}Y.  
\end{equation}
where, $j$ labels the measured qubit and $-\frac{\pi}{2}<\varphi_{j}\leq\frac{\pi}{2}$. The eigenvectors of $O_j$ are given by,
\begin{equation}
    \mathcal{B}_{j}(\varphi_{j})=\frac{|0\rangle + (-1)^{s_{j}}  e^{i\varphi_{j}} |1\rangle}{\sqrt{2}}.
    \label{eq:measurementbasis}
\end{equation}
Two measurement outcomes $s_{j}\in\{0,1\}$ are possible for each qubit and typically the measurement on the next qubit depends on the result of the measurement outcome on the previous one. Single qubit gates are implemented by measuring a row of connected qubits with suitable measurement angles $\varphi_j$. Up to four such measurements may be needed to implement an arbitrary single qubit rotation. A two qubit gate like the {\sc cnot} gate can be implemented on the two-dimensional cluster through a similar sequence of measurements on two rows of qubits \cite{raussendorf_measurement-based_2003}. In short, through the choice of suitable measurements a complete set of universal quantum gates can be realized within MBQC. Measurements on an $n$-qubit cluster state reproduces the action of a gate-based quantum computation on a register of $d$ qubits in the sense that after $n-d$ measurements, the state of the remaining qubits of the cluster will match the output of the gate-based circuit up to irrelevant overall phases. Measurement of these $d$ output qubits in a standard basis produces the measurement statistics which form the (classical) output of the computation. 

\section{Non-Classical Correlations in MBQC using Cluster States \label{flow}}

\begin{figure}[!htb]
 	\resizebox{6.7 cm}{2.1 cm}{\includegraphics{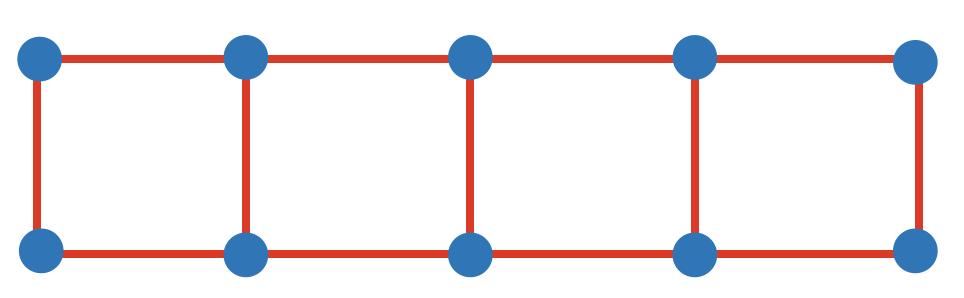}}
	\caption{A ladder-shaped cluster state is shown. At each of the vertices denoted by blue dots, a qubit is placed and initialized in the state $|+\rangle$. $CZ$ gates are applied on all pairs of Qubits linked by an edge of the graph. The two qubits on the last rung of the ladder on the right end are treated as the {\em output} qubits. The dynamics of non-classical correlations between these two output qubits is studied as measurements are performed on all qubits except the last three. \label{cluster}}
 \end{figure}

The output qubits have to be driven to a specific state with its own distinctive quantum and classical correlations by the sequence of measurements in the MBQC model. Our interest is in how such correlations develop between the output qubits. For investigating this problem we consider a simplified cluster state defined on a square lattice with a ladder shape as shown in Fig.~\ref{cluster}. The two qubits on the right end of the ladder are the output qubits.  When the ladder shaped cluster is in the initial stabilizer state, non-classical correlations are not present in the state of the two output qubits~\cite{SubsystemDiscord}. The question of interest is how such correlations develop as measurements are done on other qubits in the cluster leading to the localisation of the relevant information on these two qubits as the output of the computation. Do such correlations develop only when the immediate neighbours are measured or do they develop earlier indicating a {\em flow} of the quantum information delocalized across the cluster into these qubits at each stage? What is the nature of the non-classical correlations developed on the output qubits and how does it depend on the details of the measurements performed on the rest of the cluster. These are some of the questions that we investigate in the following. 

We point out again that the last {\em three} qubits are left unmeasured in our numerical and analytical investigations. If only the last two are left unmeasured then their joint state must be pure and the only type of non-classical correlations that can be present between them is entanglement. To investigate a richer array of phenomena, we look at the non-classical correlations in the last two qubits while leaving a third one also unmeasured. The measurement operators on the $n-3$ qubits given by Eq.~\eqref{measurementOP} are non-Clifford for all values of $\phi_j$ except $k\pi/2$, for $k=0,1,2,3$. We apply these non-Clifford gates to $n-3$ qubits of an $n$-qubit ladder state either randomly or in a systematic manner and look at non-classical correlations in the last two qubits. The length of the ladder we consider is limited by the computational resources available. 
\begin{figure}[t]
\includegraphics[width=0.95\linewidth]{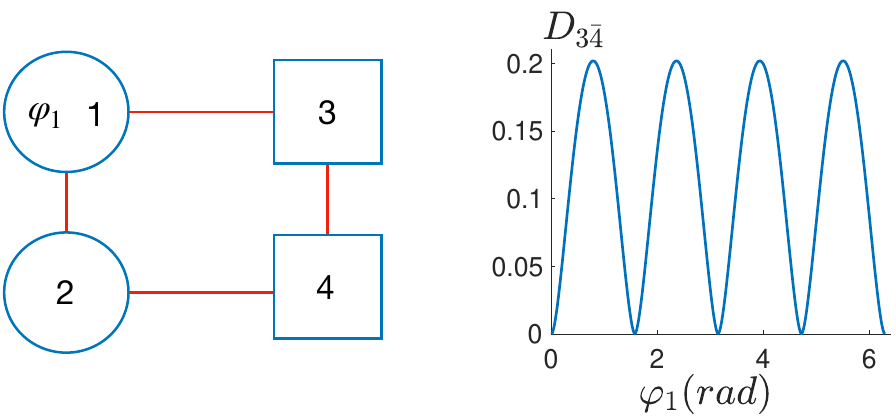}
\caption{The four-qubit ladder-type cluster state is shown on the left-hand side. Projective measurements on to the eigenstates of $O_1(\varphi_1)$ are performed on qubit 1. The quantum discord, $D_{3\bar{4}}$, computed between the third and fourth qubits as a function of angle $\varphi_{1}$ and averaged over the two possible measurement outcomes on qubit 1 is shown on the right-hand side. $D_{\bar{3}4}=0$ for all values of $\varphi_1$.   
\label{4qubit}}
\end{figure}
We first consider the simplest case of a four qubit cluster states as shown in the left-hand side of Fig.~\ref{4qubit}. A measurement on to the basis of $O_1(\varphi_1)$ is performed on qubit 1 while leaving the remaining three qubits unmeasured. We are interested in the state $\rho_{34}$ of qubits 3 and 4. Prior to the measurement on qubit 1, $\rho_{34} = \openone_4/4$, where $\openone_4$ is the identity matrix in four dimensions. After qubit 1 is measured in the basis $\mathcal{B}_{1}(\varphi_{1})$ given in Eq.~\eqref{eq:measurementbasis} with measurement angle $\varphi_{1}$, we obtain 
\begin{equation}
  \rho_{34}^{\rm out}= \frac{1}{4}\left( \begin{array}{cccc} 
	1 & A_{\varphi_{1}} & 0 & B_{\varphi_{1}} \\ 
	A_{\varphi_{1}}^{*} & 1 & B_{\varphi_{1}}^{*} & 0  \\ 
	0 & B_{\varphi_{1}} & 1 & A_{\varphi_{1}}  \\ 
	B_{\varphi_{1}}^{*} & 0 & A_{\varphi_{1}}^{*} & 1 \end{array} \right)
   \label{eq:densitymatrix}
 \end{equation}
 Where, $A_{\varphi_{1}}=(-1)^{s_{1}}\cos{\varphi_{1}},$  $B_{\varphi_{1}}=i(-1)^{s_{1}}\sin{\varphi_{1}},$ and $s_{1} \in \{0, 1\}$ depending on the result of the measurement on qubit 1. 

We quantify the non-classical correlations in $\rho_{34}^{\rm out}$ using quantum discord~\cite{Qd_zurek, Qd_vedral, Qd_mid, Qd_measures, Qd_allies}. Quantum discord is essentially the difference between the total correlations and the classical correlations in a bipartite quantum state $\rho_{\rm AB}$ of two systems $A$ and $B$. It is an entropic measure and the total correlations are quantified in terms of the mutual information $I(A:B) = S(A) + S(B) - S(AB)$ between $A$ and $B$ with $S(X)= - {\rm tr}(\rho_X \log \rho_X)$ denoting the vonNeumann entropy. Classical correlations are quantified in terms of the conditional entropy $S(A|\rho_B^j)$ where $\rho_B^j$ is the post-measurement state of $B$ obtained when a measurement indexed by $j$ is performed on subsystem $B$. In terms of this, an alternate expression for the mutual information can be written down as $J(A:B) = S(A) - \sum_j S(A|\rho_B^j)$, where the sum over $j$ ensures a complete set of measurements on $B$. If $A$ and $B$ are classical systems with the vonNeumann entropy replaced by the corresponding Shannon entropy, then by Bayes theorem, the two definitions of classical mutual information $I$ and $J$ are equal to each other. In the quantum case the two need not be equal. Defining ${\mathcal J}(A:\bar{B}) = \max_{\{\Pi_j\}} J(A|B)$, where the maximum is over all possible measurements $\{\Pi_j\}$ on $B$, quantum discord is defined as the difference,
\begin{equation}
    \label{discord1}
    D_{A\bar{B}} = I(A:B) - {\mathcal J}(A:\bar{B}),
\end{equation}
where the bar over $B$ indicates that $B$ is the system that is treated as observed/measured for computing the classical correlations between $A$ and $B$. Note that, in practice, $J(A|B)$ is maximised over all projective measurements on $B$. Quantum discord is asymmetric between the two subsystems as can be seen from its definition and in general $D_{A\bar{B}} \neq D_{\bar{A}B}$

Quantum discord $D_{3\bar{4}}$ between qubits 3 and 4 is plotted for $\varphi_{1}$ ranging from $0$ to $2\pi$ on the right-hand side of the Fig.~\ref{4qubit}. Note that $D_{\bar{3}4}$ is found to be zero for all values of $\varphi_1$ up to the uncertainties that appear in the numerical computation of the quantum Discord.  There is no entanglement between the two qubits either and concurrence between them is identically zero. We see that for $\varphi_1 =0, \pi/2, \pi, 3\pi/2$ and $2 \pi$, $D_{3\bar{4}}$ also vanishes. This can be understood easily by noting that for these values of $\varphi_1$, $O_1(\varphi_1)$ is either $X_1$ or $Y_1$. As noted earlier, a Pauli (Clifford) measurement on a member of a cluster state not only removes it from the cluster but it also ensures that the cluster state formed by the remaining qubits is a stabilizer state if the initial state is also a stabilizer state. So, for these values of $\varphi_1$ the post-measurement state of qubits $2$, $3$ and $4$ is again a stabilizer state with zero non-classical correlations in $\rho_{34}^{\rm out}$.

We can understand the points where $D_{3\bar{4}}=0$ by looking at how the stabilizer generators of qubits $2$, $3$ and $4$ transform when the measurement is performed on qubit 1. Ignoring qubit 1, prior to the measurement the three relevant generators are $g_4(2) = X_2I_3Z_4$, $g_4(3) = I_2X_3Z_4$ and $g_4(4) = Z_2Z_3X_4$, where the subscript $4$ on $g$ indicates that we are considering the $4$ qubit ladder. When qubit 1 is measured in the eigenbasis of $O_1(\varphi_1)$, these generators transform to, 
\begin{eqnarray}
\label{newgens4}
g_{4}(2)&= & X_2X_3I_4, \nonumber \\
g_{4}(3)&= & (-1)^{s_{1}} \bigl[\cos{\varphi_{1}}Z_2Z_3I_4 + \sin{\varphi_{1}}Z_2Y_3Z_4\bigr], \nonumber \\
g_{4}(4)&= & Z_2Z_3X_4. 
\end{eqnarray}
We see that the two generators $g_{4}(2)$ and $g_{4}(4)$ remain Pauli strings, but the generator $g_{4}(3)$ is not a Pauli string anymore but is a sum of two Pauli strings instead. In other words, the state of qubits $2$, $3$ and $4$ is no longer a stabilizer state. Since qubit $4$ is not directly connected to the measured qubit, the generator $g_4(4)$ associated with it remains invariant. At $\varphi_{1}=0, \pi/2, 3\pi/2,$ and $2\pi$, the generator $g_{4}(3)$ also becomes a single Pauli string and for these values of $\varphi_{1}$, the quantum discord $D_{3\bar{4}}$ is absent. 

\subsection{Six and eight qubit ladders \label{sixeight}}

\begin{figure}[!htb]
    \includegraphics[width=0.50\linewidth]{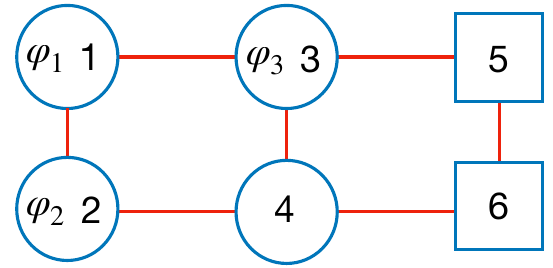}
    \caption{The six qubit ladder state with measurements done on qubits $1$, $2$ and $3$ with randomly chosen measurement angles $(\varphi_1, \varphi_2, \varphi_3)$. 
    \label{6qubit}}
\end{figure}

\begin{figure}[!htb]
     \includegraphics[width=0.93\linewidth]{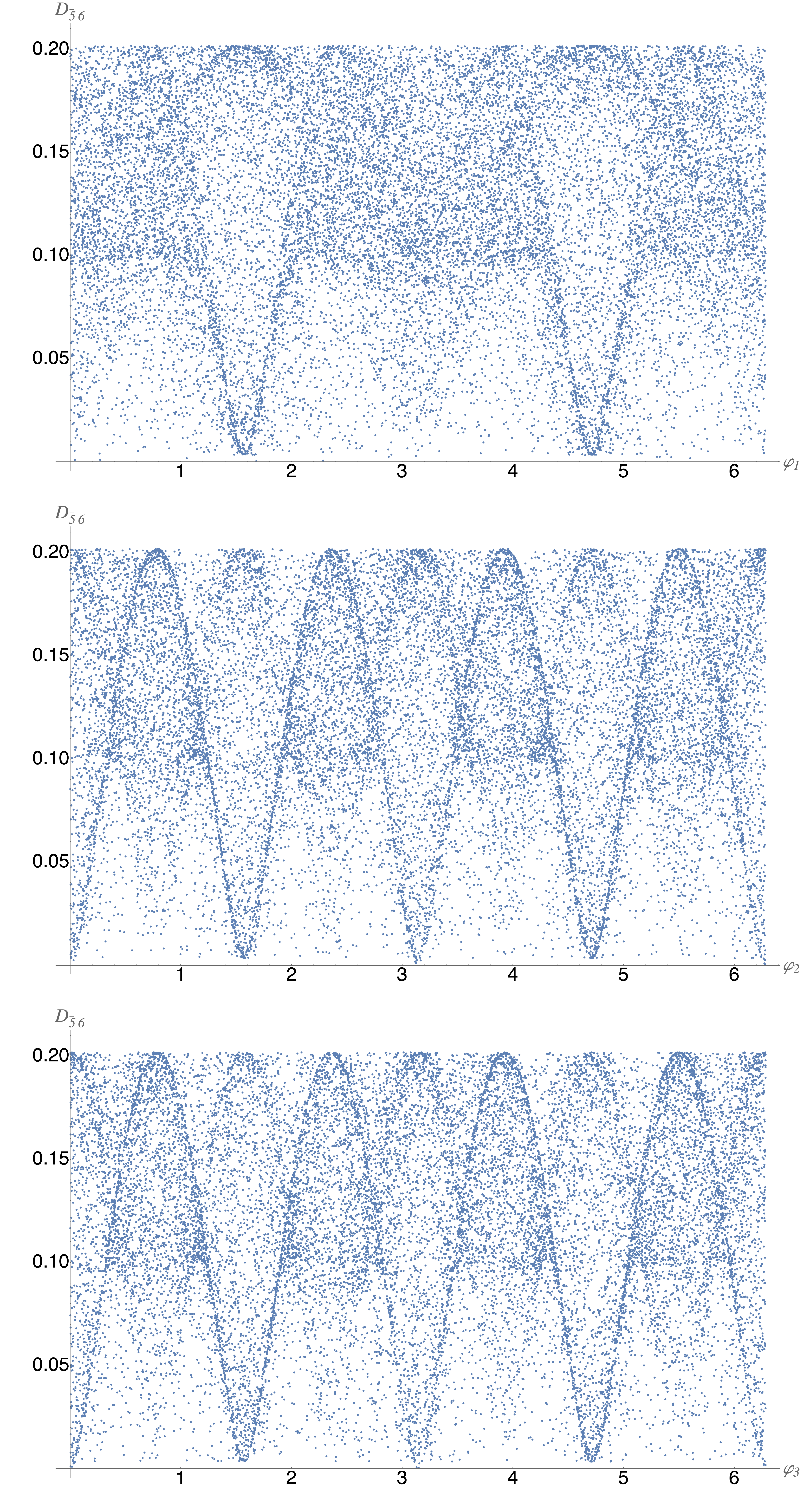}
    \caption{The top panel shows $D_{\bar{5}6}$ as a function of $\varphi_1$ for 20,000 random choices of $(\varphi_1, \varphi_2, \varphi_3)$ for the six qubit ladder state. The middle and bottom panel shows $D_{\bar{5}6}$ as function of $\varphi_2$ and $\varphi_3$ respectively. $D_{\bar{5}6}$ for each choice of $(\varphi_1, \varphi_2, \varphi_3)$ are averaged over the eight possible measurement outcomes $(s_j = \pm 1)$. 
    \label{6qubita}}
\end{figure}

We now consider the six qubit ladder cluster state shown in Fig.~\ref{6qubit}.
Measurements are done on all except the last three qubits in the basis ${\mathcal{B}}_{j}(\varphi_{j})$ with randomly chosen measurement angles $\varphi_j$. The quantum discords $D_{5\bar{6}}$ and $D_{\bar{5}6}$ were computed for 20,000 combinations of $\varphi_1$, $\varphi_2$, $\varphi_3$ with the discords being averaged over the eight possible combinations of measurement results $(s_j = \pm 1)$ on qubits $1$, $2$ and $3$. We find that in this case $D_{5\bar{6}}$ is zero up to the uncertainties that appear in the numerical computation of quantum Discord. We see that this is different from the pattern for the 4-qubit case where $D_{\bar{3}4}$ was zero. The non-zero quantum discord, $D_{\bar{5}6}$, is plotted as a function of each of the three measurement angles $\varphi_1$, $\varphi_2$ and $\varphi_3$ in Fig.~\ref{6qubita}. There is no entanglement in $\rho_{56}^{\rm out}$ for any of the values of the measurement angles chosen. 

We can track the flow of non-classical correlations in the six qubit ladder by computing  $D_{\bar{5}6}$ when measurements are done one by one on qubits $1$, $2$ and $3$. When the  measurements are on the first and second qubits only then both $D_{5\bar{6}}$ and $D_{\bar{5}6}$ are zero. This is because the measurement operators commute with the CZ gates applied between qubits (3,5), (4,6), and (5,6), thereby leaving the output state (5,6) unaffected. However, the effect of the measurements on $1$ and $2$ do eventually flow to qubits $5$ and $6$ when qubit $3$ is measured as can be seen from Fig.~\ref{6qubita}. As a function of the last measurement angle $\varphi_3$ shown in the bottom panel of Fig.~\ref{6qubita}, $D_{\bar{5}6}$ shows a discernible pattern with two distinct oscillatory patterns. As a function of the second-last measurement angle $\varphi_2$, the same pattern persists but is a bit less well defined. The top panel of Fig.~\ref{6qubita} shows that even if there is a pattern in the relationship between $D_{\bar{5}6}$ and the first measurement angle $\varphi_1$, it is masked by the subsequent measurements and does not show up as clearly as in the other case. Note that the maximum values for $D_{\bar{5}6}$ for the six qubit case is same as the maximum for $D_{3\bar{4}}$ for four qubits. The stabilizer generators $g_{6}(4)$, $g_{6}(5)$ and $g_{6}(6)$ transform under the measurements specified by $(\varphi_1, \varphi_2, \varphi_3)$ as
\begin{eqnarray}
g_{6}(4) \! &= & \!\bigl[\cos{\varphi_{3}}(\cos{\varphi_{2}}I_4Z_5I_6 +\sin{\varphi_{2}}X_4Y_5I_6) \nonumber  \\ & & \quad  + \sin{\varphi_{3}}(\cos{\varphi_{2}}I_4Y_5Z_6-\sin{\varphi_{2}}X_4Z_5Z_6)\bigr], \nonumber \\  
g_{6}(5) \!&=& \! \bigl[-\sin{\varphi_{1}}\sin{\varphi_{3}}Y_4Y_5I_6 + (-1)^{s_{1}}\cos{\varphi_{1}}X_4I_5Z_6 \nonumber \\&& \qquad \qquad  - \sin{\varphi_{1}}\cos{\varphi_{3}}Y_4Z_5Z_6\bigr], \nonumber \\
g_{6}(6) \! &=& \! Z_4Z_5X_6,  
\end{eqnarray}
The periodicity with respect to the three angles $\varphi_1$, $\varphi_2$ and $\varphi_3$ seen in Fig.~\ref{6qubita} is also reflected in the stabiliser generators. We see that the measurement on qubit 1 affects $g_6(5)$  through the measurement on 3 while this measurement with angle $\varphi_1$ has no effect on $g_6(4)$ explaining the simpler periodicity of the top panel of Fig.~\ref{6qubita}. As in the 4-qubit case, the generator $g_6(6)$ associated with qubit 6 remains unchanged since none of the qubits directly connected to qubit $6$ are measured.

\begin{figure}[!htb]
    \includegraphics[width=0.70\linewidth]{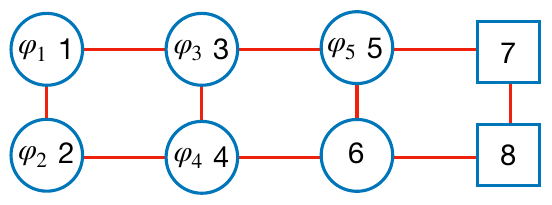}
    \caption{The eight qubit ladder state with measurements done on qubits $1$, $2$, $3$, $4$ and $5$ with randomly chosen measurement angles $(\varphi_1, \varphi_2, \varphi_3, \varphi_4, \varphi_5)$. 
    \label{8qubit}}
\end{figure}

We also investigated numerically the eight-qubit ladder shown in Fig.~\ref{8qubit}, computing the average discord between qubits $7$ and $8$. 

\begin{figure}[b]
    \includegraphics[width=0.7\linewidth]{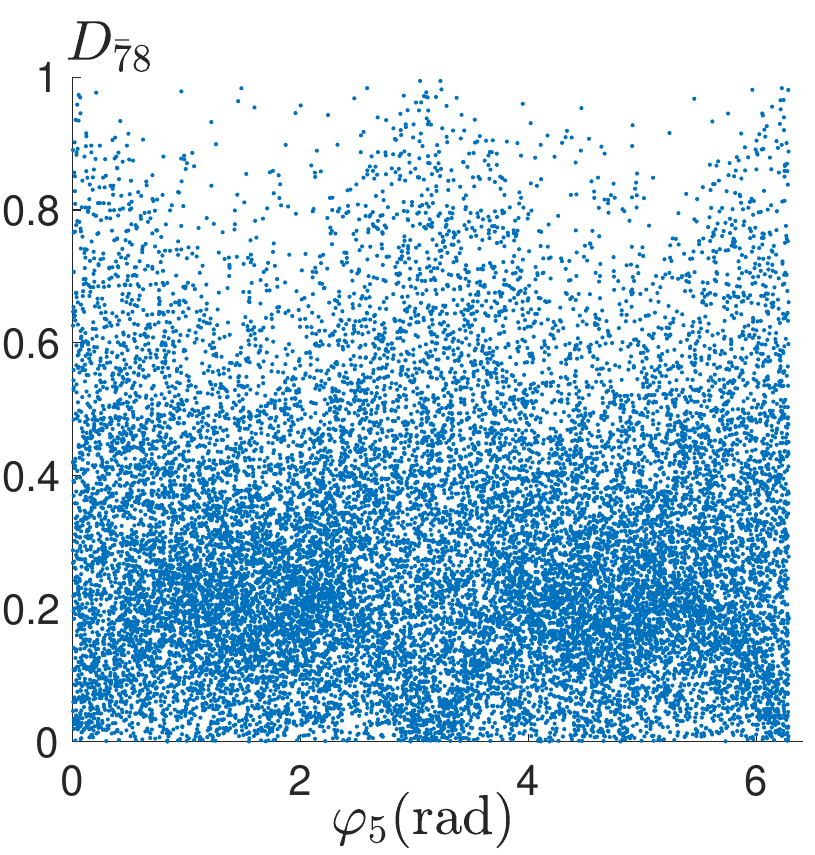}
    \includegraphics[width=0.7\linewidth]{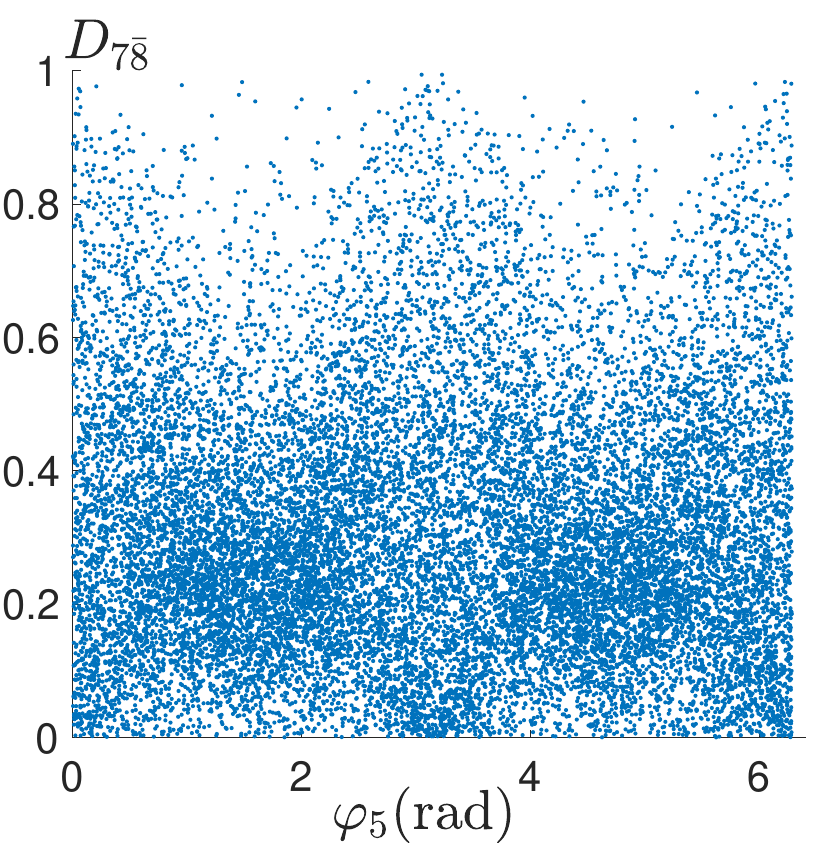}
    \caption{ Quantum discord $D_{\bar{7}8}$  averaged over all possible measurement outcomes on the proceeding 5 qubits plotted against the angle $\varphi_5$ of the measurement on qubit $5$ for 20,000 randomly chosen values of the measurement angles ($\varphi_{1}, \varphi_{2}, \varphi_{3}, \varphi_{4}, \varphi_{5}$) is shown in the top panel. The bottom panel shows $D_{7\bar{8}}$ as a function of $\varphi_5$.
    \label{8qubita}}
\end{figure}
\begin{figure}[!htb]
    \includegraphics[width=0.70\linewidth]{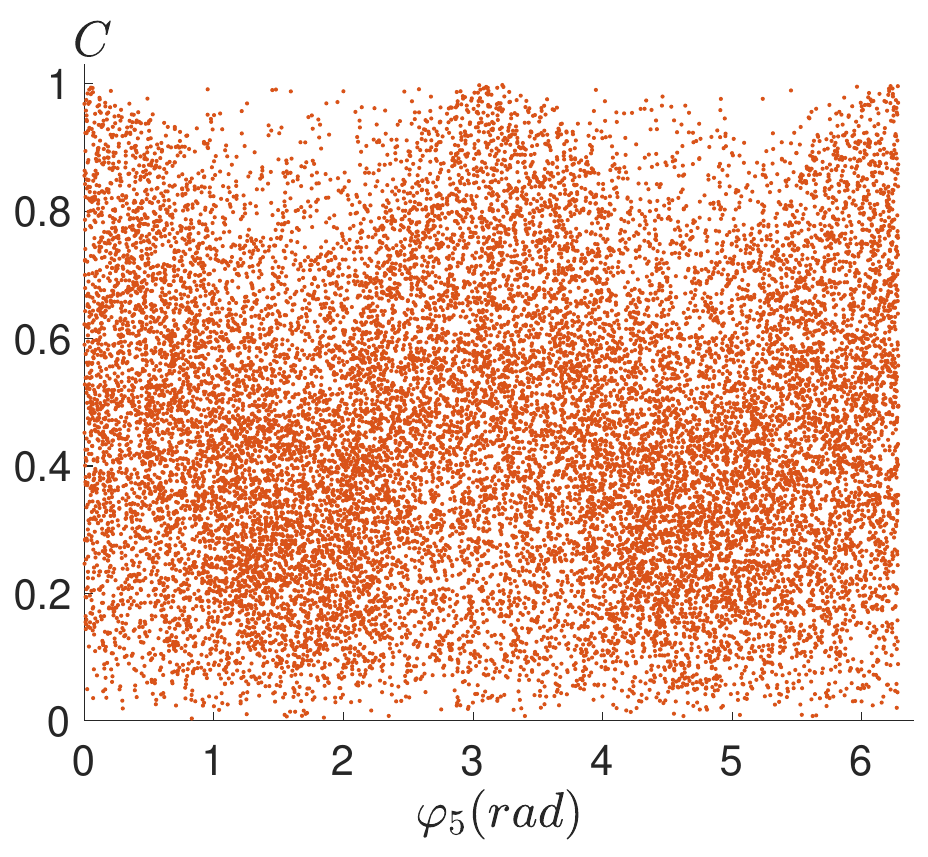}
    \caption{The concurrence, $C$ between the last two qubits of the 8 qubit ladder averaged over all possible measurement outcomes on the proceeding 5 qubits are plotted against the angle $\varphi_5$ for 20,000 randomly chosen values of the measurement angles.
    \label{8qubitb}}
\end{figure}

Quantum discords $D_{7\bar{8}}$ and $D_{\bar{7}8}$ averaged over the $2^5=32$ possible measurement outcomes on the proceeding 5 qubits are plotted in Fig.~\ref{8qubita} against the angle $\varphi_5$ of the measurement on qubit $5$ which is directly connected to qubit $7$ of the cluster. The discord is computed again for 20,000 randomly chosen values of the measurement angles $\varphi_i$, $i=1,\ldots,5$. We see that with the longer ladder, the non-classical correlations developed between the qubits in the last rung is stronger with both $D_{7\bar{8}}$ and $D_{\bar{7}8}$ having a maximum value of $1$. The asymmetry between the two discords also is not present indicating the presence of entanglement between the two. Entanglement as quantified by the concurrence was also computed for the 20,000 choices of measurement angles and the result is plotted in Fig.~\ref{8qubitb} showing that indeed the qubits in the last rung of the ladder does get entangled when the ladder is sufficiently long.

\subsection{Effect of the last measurement \label{last}}

We now focus on the effect of the last measurement on the behaviour of non-classical correlations in the qubits on the last run. In our case this measurement is on the $(n-3)^{\rm rd}$ qubit. We keep the values of $\varphi_i$, $i=1,\ldots, n-4$ fixed at values that are chosen at random and vary on $\varphi_{n-3}$. Here we compute $D_{n-1 \, \bar{n}}$ and $D_{\overline{n-1} \, n}$ as well as the concurrence between the last two qubits for $4$, $6$, $8$ and $10$ qubit ladders. In each case the discord is averaged over all possible measurement outcomes for the preceding measurements. For the first example we consider, the fixed measurement angles chosen for $\varphi_i$, $i=1,\ldots,n-4$ (with $\varphi_{n-3}$ varied in the interval $0$ to $2\pi$) are given in Table~\ref{table1}:
\begin{center}
    \begin{table}[!htb]
    \begin{tabular}{|l|c|c|c|c|c|c|}
      \hline
        & $\varphi_1$ & $\varphi_2$ & $\varphi_3$ & $\varphi_4$ & $\varphi_5$ & $\varphi_6$ \\
      \hline
      4-qubit   & - & - & - & - & - & -   \\
      6-qubit   & $0.37\pi$ & $0.78\pi$  & - & - & - & -  \\
      8-qubit   & $1.5\pi$ & $0.19\pi$ & $0.37\pi$ & $0.78\pi$ & - & - \\
      10-qubit  & $0.24\pi$ & $0.53\pi$ & $1.5\pi$ &  $0.19\pi$ & $0.37\pi$ & $0.78\pi$ \\
      \hline
    \end{tabular}
    \caption{Fixed measurement angles for Example 1 \label{table1}}
    \end{table}
\end{center}

\begin{figure}[!htb] 
    \includegraphics[width=0.67\linewidth]{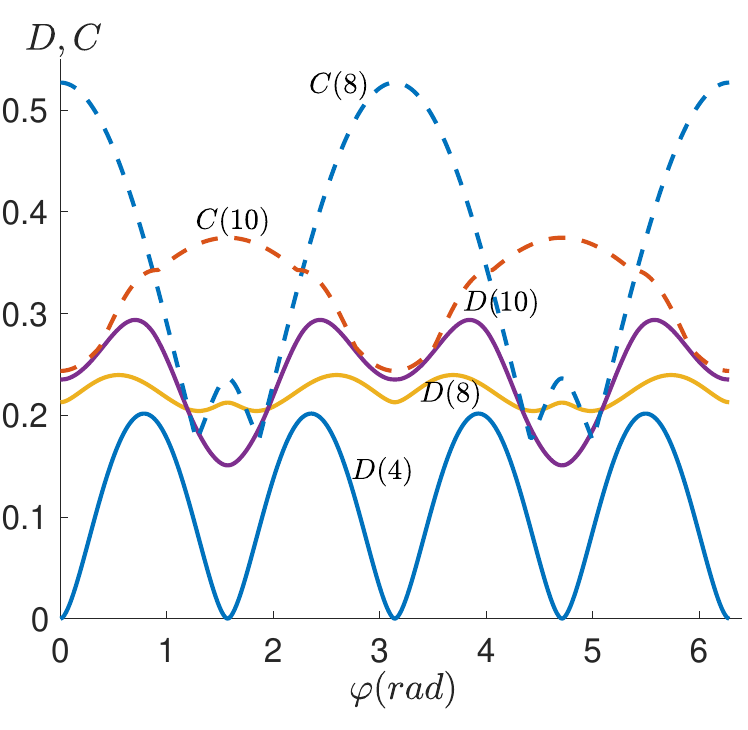}
    \includegraphics[width=0.70\linewidth]{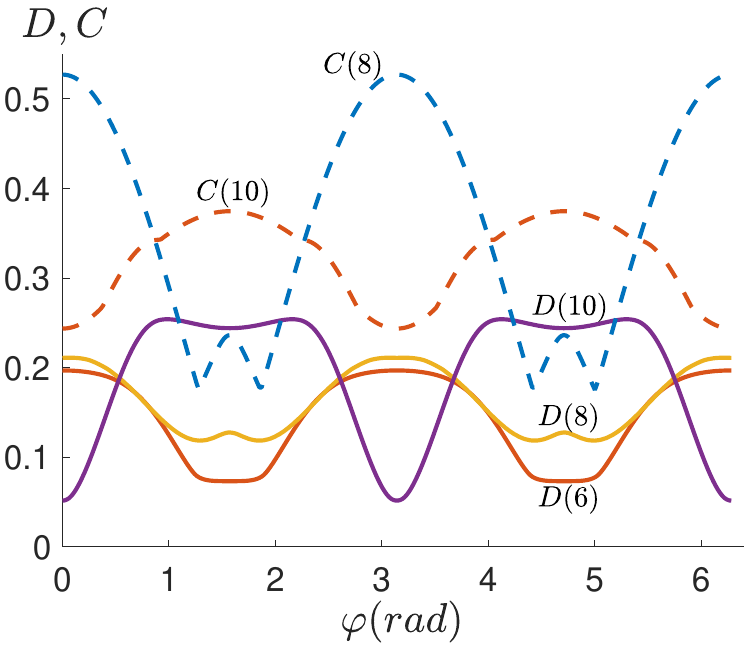}
    \caption{The plot on top shows $D_{n-1 \, \bar{n}}$  vs $\varphi_{n-3}$ while the plot at the bottom shows $D_{\overline{n-1} \, n}$ vs $\varphi_{n-3}$ for 4, 6, 8 and 10 qubits. Note that $D_{\bar{3}4}$ and $D_{5\bar{6}}$ are zero for the 4 and 6 qubit ladders respectively. The fixed values used for all measurement angles except $\varphi_{n-3}$ are listed in Table~\ref{table1}.The corresponding concurrence is also shown with dotted lines in both plots. \label{fixedSet1}}
\end{figure}

For the choice of fixed measurement angles in Table~\ref{table1} we plotted the two discords and concurrence as a function of $\varphi_{n-3}$ in Fig.~\ref{fixedSet1}. We see from the plots in Fig.~\ref{fixedSet1} that both $D_{n-1 \, \bar{n}}$ and $D_{\overline{n-1} \, n}$ show an increasing trend with increasing number of qubits in the ladder. Furthermore, for the 8 and 10 qubit cases, the last two qubits become entangled as well and the non-classical correlations is progressively dominated by entanglement. The behaviour is not strictly monotonic but rather it is so on an average. For instance, $D_{5\bar{6}}$ is zero in the six qubit case for all values of $\varphi_3$ even if $D_{3\bar{4}}$ is non-zero for the corresponding 4 qubit case. The concurrence does not show a specific pattern for the two cases (8 and 10 qubits) for which it is non-zero and computations involving more qubits which are prohibitively expensive in terms of compute time will be required to see any trend if it exists.   

\begin{figure}[!htb]
    \includegraphics[width=0.70\linewidth]{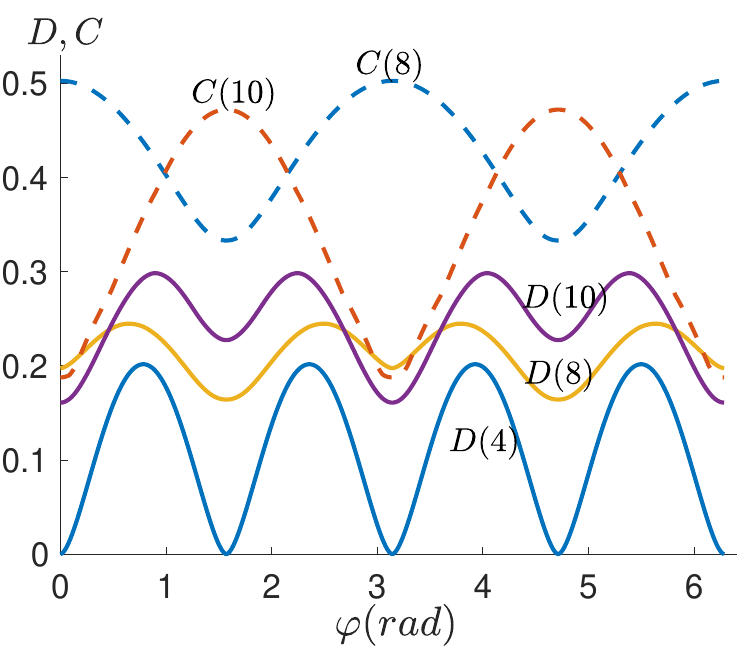}
    \includegraphics[width=0.70\linewidth]{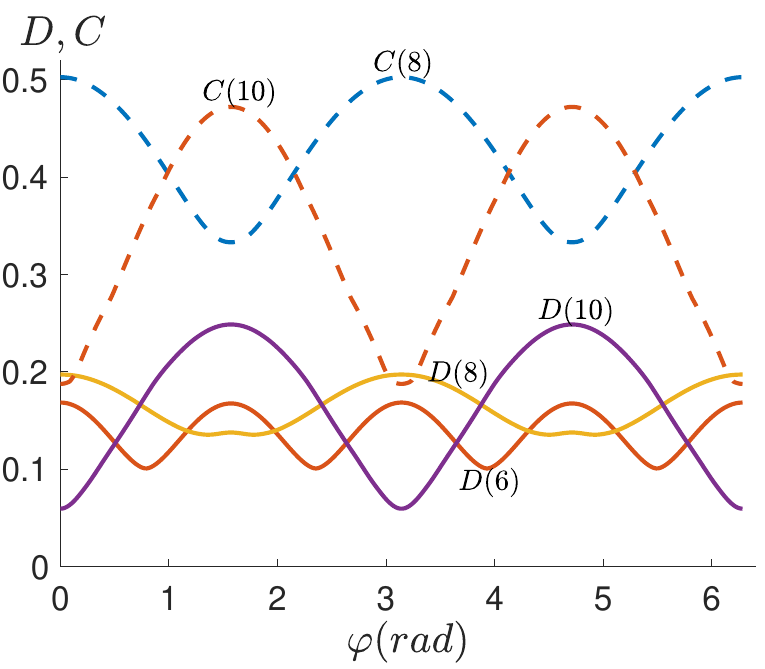}
    \caption{The plot on top shows $D_{n-1 \, \bar{n}}$  vs $\varphi_{n-3}$ while the plot at the bottom shows $D_{\overline{n-1} \, n}$ vs $\varphi_{n-3}$ for 4, 6, 8 and 10 qubits. The fixed values used for all measurement angles except $\varphi_{n-3}$ are listed in Table~\ref{table2}. The corresponding concurrence is also shown with dotted lines in both plots. \label{fixedSet2}}
\end{figure}

The values for the measurement angles chosen for the second example we investigated is given in Table~\ref{table2}:
\begin{center}
 \begin{table}[!htb]
    \begin{tabular}{|l|c|c|c|c|c|c|}
      \hline
        & $\varphi_1$ & $\varphi_2$ & $\varphi_3$ & $\varphi_4$ & $\varphi_5$ & $\varphi_6$ \\
      \hline
      4-qubit   & - & - & - & - & - & -   \\
      6-qubit   & $1.3\pi$ & $0.67\pi$  & - & - & - & -  \\
      8-qubit   & $0.95\pi$ & $1.7\pi$ & $1.3\pi$ & $0.67\pi$ & - & - \\
      10-qubit  & $0.59\pi$ & $1.13\pi$ & $0.95\pi$ &  $1.7\pi$ & $1.3\pi$ & $0.67\pi$ \\
      \hline
    \end{tabular}
    \caption{Fixed measurement angles for Example 2 \label{table2}}
    \end{table}
\end{center}
The corresponding graphs for $D_{n-1 \, \bar{n}}$, $D_{\overline{n-1} \, n}$ and the concurrence as a function of $\varphi_{n-3}$ are shown in Fig.~\ref{fixedSet2}. Again a trend similar to that in the first example is seen. 

\subsection{A single non-Clifford measurement \label{nonCliffSingle}}

We look at the dependence of quantum Discord between the qubits on the last rung of the ladder on the location of a specific non-Clifford measurement on one of the remaining $n-3$ qubits. We move this measurements from the first qubit on the first rung of the ladder, step-by-step till the $(n-3)^{\rm rd}$ qubit while performing Clifford measurements on the remaining $n-4$ qubits. Specifically the Clifford measurements done on the $n-4$ qubits are either along the $X$ or $Y$ bases. The results we obtain for ladders with $8$, $10$, $12$ and $14$ qubits respectively are summarized in Fig.~\ref{OneNonCliff}. in the figure, the Clifford measurements are along the $X$ basis while the non-Clifford measurement on one of the qubits is along $O(\pi/3)$ which corresponds to measurements along $(|0\rangle \pm e^{1\pi/3}|1\rangle)/\sqrt{2}$. We see that the dependence of the quantum discord in the last rung (qubits labelled as $A$ and $B$ and shown in blue in the figure) on the location of the non-Clifford measurement is quite non-intuitive. The locations where the non-Clifford measurement leads to discord between the qubits on the last rung are marked in red in Fig.~\ref{OneNonCliff}. The values of the discord produced on the last rung are marked near the respective qubits where the non-Clifford measurement is done. The  numbers shown are $D^{\rm avg}_{A\bar{B}}$ on the top right, $D^{\rm avg}_{\bar{A}B }$ on the bottom right and the concurrence $C_{AB}$, whenever it is present, on the bottom left of each of the measured qubits. $D^{\rm avg}$ denotes the average value of discord computed over the $2^{n-3}$ possible combinations of ``plus" and ``minus" measurements on the measured qubits. Non-Clifford measurements on qubits at many locations do not generate any non-classical correlations between $A$ and $B$. These qubits are shown in gray in Fig.~\ref{OneNonCliff} and discord values are not listed around the corresponding qubits. It turns out that the values of  $D_{A\bar{B}}$ and $D_{\bar{A}B}$ do not depend on whether a ``plus" or ``minus" projection is done at each site and therefore $D^{\rm avg}_{A\bar{B}}$ and $D^{\rm avg}_{\bar{A}B}$ are just equal to  $D_{A\bar{B}}$ and $D_{\bar{A}B}$ for any choice. 

\begin{figure}[!htb]
	\resizebox{\columnwidth}{!}
	{
		\begin{tikzpicture}
			
			\draw[gray, very thick](-5, 6.0) -- (-3, 6.0);
			\draw[gray, very thick](-5, 6.0) -- (-5, 8.0);
			\draw[gray, very thick](-5, 8.0) -- (-3, 8.0);
			
			\draw[gray, very thick](-3, 6.0) -- (-1, 6.0);
			\draw[gray, very thick](-3, 6.0) -- (-3, 8.0);
			\draw[gray, very thick](-3, 8.0) -- (-1, 8.0);
			
			\draw[gray, very thick](-1, 6.0) -- (1, 6.0);
			\draw[gray, very thick](-1, 6.0) -- (-1, 8.0);
			\draw[gray, very thick](-1, 8.0) -- (1, 8.0);
			
			\draw[gray, very thick](1, 6.0) -- (1, 8.0);

			\draw[gray,fill=red] (-5, 6.0) circle (0.2) 
                node[anchor=south west]{\textcolor{blue}{\bfseries $\;\; 1.0$}} 
				node[anchor=north west]{\textcolor{black}{\bfseries $\;\; 1.0$}}
                node[anchor=north east]{\textcolor{teal}{\bfseries $ 1.0 \;$}};
			
			\draw[gray,fill=red] (-3, 6.0) circle (0.25)
				node[anchor=south west]{\textcolor{blue}{\bfseries $\;\; 0.1887$}} 
				node[anchor=north west]{\textcolor{black}{\bfseries $\;\; 0.1887$}} 
                node[anchor=north east]{\textcolor{teal}{\bfseries $ 0.5 \;$}};
			
			\draw[gray,fill=violet] (-1, 6.0) circle (0.25);

			\draw[gray,fill=blue] (1, 6.0) circle (0.25);
											
			\draw[gray,fill=red] (-5, 8.0) circle (0.25)
				node[anchor=south west]{\textcolor{blue}{\bfseries $\;\; 0.1887$}} 
				node[anchor=north west]{\textcolor{black}{\bfseries $\;\; 0.1887$}}
                node[anchor=north east]{\textcolor{teal}{\bfseries $ 0.5 \;$}};
			
			\draw[gray,fill=red] (-3, 8.0) circle (0.25)
                node[anchor=south west]{\textcolor{blue}{\bfseries $\;\; 1.0$}} 
				node[anchor=north west]{\textcolor{black}{\bfseries $\;\; 1.0$}}
                node[anchor=north east]{\textcolor{teal}{\bfseries $ 1.0 \;$}};
			
			\draw[gray,fill=red] (-1, 8.0) circle (0.25)
                node[anchor=south west]{\textcolor{blue}{\bfseries $\;\; 1.0$}} 
				node[anchor=north west]{\textcolor{black}{\bfseries $\;\; 1.0$}}
                node[anchor=north east]{\textcolor{teal}{\bfseries $ 1.0 \;$}};
			
			\draw[gray,fill=blue] (1, 8.0) circle (0.25);

			\node at (-5,8.0) {\textcolor{white}{1}};
			\node at (-5,6.0) {\textcolor{white}{2}};
			\node at (-3,8.0) {\textcolor{white}{3}};
			\node at (-3,6.0) {\textcolor{white}{4}};
			\node at (-1,8.0) {\textcolor{white}{5}};
			\node at (-1,6.0) {\textcolor{white}{C}};
			\node at (1,8.0) {\textcolor{white}{A}};
			\node at (1,6.0) {\textcolor{white}{B}};

			
			\draw[gray, very thick](-5, 2.5) -- (-3, 2.5);
			\draw[gray, very thick](-5, 2.5) -- (-5, 4.5);
			\draw[gray, very thick](-5, 4.5) -- (-3, 4.5);
			
			\draw[gray, very thick](-3, 2.5) -- (-1, 2.5);
			\draw[gray, very thick](-3, 2.5) -- (-3, 4.5);
			\draw[gray, very thick](-3, 4.5) -- (-1, 4.5);
			
			\draw[gray, very thick](-1, 2.5) -- (1, 2.5);
			\draw[gray, very thick](-1, 2.5) -- (-1, 4.5);
			\draw[gray, very thick](-1, 4.5) -- (1, 4.5);
			
			\draw[gray, very thick](1, 2.5) -- (1, 4.5);
			
			\draw[gray, very thick](3, 2.5) -- (3, 4.5);
			
			\draw[gray, very thick](1, 2.5) -- (3, 2.5);
			
			\draw[gray, very thick](1, 4.5) -- (3, 4.5);
			
			\draw[gray,fill=red] (-5, 2.5) circle (0.2) 
				node[anchor=south west]{\textcolor{blue}{\bfseries $\; 0.1659$}} ;
			
			\draw[gray,fill=gray] (-3, 2.5) circle (0.25); 
			
			\draw[gray,fill=gray] (-1, 2.5) circle (0.25); 
				
			\draw[gray,fill=violet] (1, 2.5) circle (0.25);
											
			\draw[gray,fill=gray] (-5, 4.5) circle (0.25); 
			
			\draw[gray,fill=red] (-3, 4.5) circle (0.25) 
				node[anchor=south west]{\textcolor{blue}{\bfseries $\; 0.1659$}}; 
			
			\draw[gray,fill=gray] (-1, 4.5) circle (0.25); 
			
			\draw[gray,fill=gray] (1, 4.5) circle (0.25); 
			
			\draw[gray,fill=blue] (3, 4.5) circle (0.25);
			
			\draw[gray,fill=blue] (3, 2.5) circle (0.25);

			\node at (-5,4.5) {\textcolor{white}{1}};
			\node at (-5,2.5) {\textcolor{white}{2}};
			\node at (-3,4.5) {\textcolor{white}{3}};
			\node at (-3,2.5) {\textcolor{white}{4}};
			\node at (-1,4.5) {\textcolor{white}{5}};
			\node at (-1,2.5) {\textcolor{white}{6}};
			\node at (1,4.5) {\textcolor{white}{7}};
			\node at (1,2.5) {\textcolor{white}{C}};
			\node at (3,4.5) {\textcolor{white}{A}};
			\node at (3,2.5) {\textcolor{white}{B}};

			
			\draw[gray, very thick](-5, -1) -- (-3, -1);
			\draw[gray, very thick](-5, -1) -- (-5, 1);
			\draw[gray, very thick](-5, 1) -- (-3, 1);
			
			\draw[gray, very thick](-3, -1) -- (-1, -1);
			\draw[gray, very thick](-3, -1) -- (-3, 1);
			\draw[gray, very thick](-3, 1) -- (-1, 1);
			
			\draw[gray, very thick](-1, -1) -- (1, -1);
			\draw[gray, very thick](-1, -1) -- (-1, 1);
			\draw[gray, very thick](-1, 1) -- (1, 1);
			
			\draw[gray, very thick](1, -1) -- (3, -1);
			\draw[gray, very thick](1, -1) -- (1, 1);
			\draw[gray, very thick](1, 1) -- (3, 1);

			\draw[gray, very thick](3, -1) -- (5, -1);
			\draw[gray, very thick](3, -1) -- (3, 1);
			\draw[gray, very thick](3, 1) -- (5, 1);
			
			\draw[gray, very thick](5, -1) -- (5, 1);
			
			\draw[gray,fill=gray] (-5, -1) circle (0.2);
			
			\draw[gray,fill=red] (-3, -1) circle (0.25) 
				node[anchor=north west]{\textcolor{black}{\bfseries $\; 0.1659$}}; 
			
			\draw[gray,fill=gray] (-1, -1) circle (0.25); 
				
			\draw[gray,fill=red] (1, -1) circle (0.25) 
				node[anchor=north west]{\textcolor{black}{\bfseries $\; 0.1659$}}; 
				
			\draw[gray,fill=violet] (3, -1) circle (0.25);
				
			\draw[gray,fill=blue] (5, -1) circle (0.25);
			
			\draw[gray,fill=red] (-5, 1) circle (0.25) 
				node[anchor=north west]{\textcolor{black}{\bfseries $\; 0.1659$}}; 
			
			\draw[gray,fill=gray] (-3, 1) circle (0.25); 
			
			\draw[gray,fill=gray] (-1, 1) circle (0.25); 
			
			\draw[gray,fill=gray] (1, 1) circle (0.25); 
			
			\draw[gray,fill=red] (3, 1) circle (0.25) 
				node[anchor=north west]{\textcolor{black}{\bfseries $\; 0.1659$}}; 
			
			\draw[gray,fill=blue] (5, 1) circle (0.25);
			
			\node at (-5,1) {\textcolor{white}{1}};
			\node at (-5,-1) {\textcolor{white}{2}};
			\node at (-3,1) {\textcolor{white}{3}};
			\node at (-3,-1) {\textcolor{white}{4}};
			\node at (-1,1) {\textcolor{white}{5}};
			\node at (-1,-1) {\textcolor{white}{6}};
			\node at (1,1) {\textcolor{white}{7}};
			\node at (1,-1) {\textcolor{white}{8}};
			\node at (3,1) {\textcolor{white}{9}};
			\node at (3,-1) {\textcolor{white}{C}};
			\node at (5,1) {\textcolor{white}{A}};
			\node at (5,-1) {\textcolor{white}{B}};
			
			
			\draw[gray, very thick](-5, -4.5) -- (-3, -4.5);
			\draw[gray, very thick](-5, -4.5) -- (-5, -2.5);
			\draw[gray, very thick](-5, -2.5) -- (-3, -2.5);
			
			\draw[gray, very thick](-3, -4.5) -- (-1, -4.5);
			\draw[gray, very thick](-3, -4.5) -- (-3, -2.5);
			\draw[gray, very thick](-3, -2.5) -- (-1, -2.5);
			
			\draw[gray, very thick](-1, -4.5) -- (1, -4.5);
			\draw[gray, very thick](-1, -4.5) -- (-1, -2.5);
			\draw[gray, very thick](-1, -2.5) -- (1, -2.5);
			
			\draw[gray, very thick](1, -4.5) -- (3, -4.5);
			\draw[gray, very thick](1, -4.5) -- (1, -2.5);
			\draw[gray, very thick](1, -2.5) -- (3, -2.5);

			\draw[gray, very thick](3, -4.5) -- (5, -4.5);
			\draw[gray, very thick](3, -4.5) -- (3, -2.5);
			\draw[gray, very thick](3, -2.5) -- (5, -2.5);
			
			\draw[gray, very thick](5, -4.5) -- (5, -2.5);
			\draw[gray, very thick](5, -4.5) -- (7, -4.5);
			\draw[gray, very thick](5, -2.5) -- (7, -2.5);
			
			\draw[gray, very thick](7, -4.5) -- (7, -2.5);

			\draw[gray,fill=red] (-5, -4.5) circle (0.2)
				node[anchor=north west]{\textcolor{black}{\bfseries $\; 0.165$}};
			
			\draw[gray,fill=gray] (-3, -4.5) circle (0.25);
			
			\draw[gray,fill=gray] (-1, -4.5) circle (0.25);
				
			\draw[gray,fill=gray] (1, -4.5) circle (0.25);
				
			\draw[gray,fill=red] (3, -4.5) circle (0.25)
				node[anchor=north west]{\textcolor{black}{\bfseries $\; 0.1659$}} ;
				
			\draw[gray,fill=violet] (5, -4.5) circle (0.25);
			
			\draw[gray,fill=gray] (-5, -2.5) circle (0.25);
			
			\draw[gray,fill=red] (-3, -2.5) circle (0.25)
				node[anchor=north west]{\textcolor{black}{\bfseries $\; 0.1659$}};
			
			\draw[gray,fill=gray] (-1, -2.5) circle (0.25);
			
			\draw[gray,fill=red] (1, -2.5) circle (0.25)
				node[anchor=north west]{\textcolor{black}{\bfseries $\; 0.1659$}} ;
			
			\draw[gray,fill=gray] (3, -2.5) circle (0.25);
			
			\draw[gray,fill=gray] (5, -2.5) circle (0.25);
			
			\draw[gray,fill=blue] (7, -2.5) circle (0.25);
			
			\draw[gray,fill=blue] (7, -4.5) circle (0.25);
			
			\node at (-5,-2.5) {\textcolor{white}{1}};
			\node at (-5,-4.5) {\textcolor{white}{2}};
			\node at (-3,-2.5) {\textcolor{white}{3}};
			\node at (-3,-4.5) {\textcolor{white}{4}};
			\node at (-1,-2.5) {\textcolor{white}{5}};
			\node at (-1,-4.5) {\textcolor{white}{6}};
			\node at (1,-2.5) {\textcolor{white}{7}};
			\node at (1,-4.5) {\textcolor{white}{8}};
			\node at (3,-2.5) {\textcolor{white}{9}};
			\node at (3,-4.5) {\textcolor{white}{10}};
			\node at (5,-2.5) {\textcolor{white}{11}};
			\node at (5,-4.5) {\textcolor{white}{C}};
			\node at (7,-2.5) {\textcolor{white}{A}};
			\node at (7,-4.5) {\textcolor{white}{B}};

		\end{tikzpicture}

	}	
\caption{Ladders with different number of rungs are shown along with quantum discord(s) between qubits $A$ and $B$ corresponding to a single non-Clifford measurement $O(\pi/3)$ on one of the $n-3$ other qubits in each ladder. On the remaining $n-4$ qubits, measurements in the $X$ basis is performed. Marked in red are the qubits on which the non-Clifford measurements lead to non-zero discord between $A$ and $B$ (qubits shown in Blue). Non-Clifford measurements on the other qubits shown in gray do not lead to discord between qubits $A$ and $B$. Also shown near the respective qubits are the values of the average discords produced between qubits $A$ and $B$.  $D^{\rm avg}_{A\bar{B}}$ is on the top right (blue) and $D^{\rm avg}_{\bar{A}B }$ is on the bottom right (black). The concurrence $C_{AB}$, if present is listed on the bottom left (teal) The average in $D^{\rm avg}$ is over all $2^{n-3}$ combinations of ``plus" and ``minus" measurements of each of the measured qubits in their respective measurement basis. It turns out that the values of discords are independent of these combinations and they depend only on the location and angle of the single non-Clifford measurement. \label{OneNonCliff} }
\end{figure}

We see that there is no discernible pattern for the locations at which non-Clifford measurements lead to discord between the qubits in the last rung. We however notice that the value of $D_{A\bar{B}}$ and $D_{\bar{A}B}$ are independent of the location of the measurement whenever non-zero discord is produced and it depends only on the choice of angle for the non-Clifford measurement. For single non-Clifford measurement corresponding to angles other than $\pi/3$ identical patterns are seen with only the values of $D^{\rm avg}$ changing according to the angle. A similar pattern is seen if the $X$-basis measurements on the remaining $n-4$ qubits are replaced by $Y$-basis measurements. There is no specific pattern available that predicts the location of the non-Clifford measurement that would lead to a finite value of discord. The locations seem quite randomly placed and independent of the number of rungs in the ladder. The ladder with 10 qubits appears to be an exception in that non-Clifford measurements on any one of $n-3$ qubits leads to not only non-zero discord but also non-zero concurrence with the discord and concurrence attaining their maximum possible values in several cases. This again serves to highlight the non intuitive manner in which quantum information distributes itself across the qubits in the ladder and how it is transformed under measurement on all but the last three qubits.

\section{Conclusion \label{conclusion}}

We have explored the non-classical correlations in two-qubit subsystems of ladder-like cluster states when measurements are done on all but the last three qubits. Our results show how projective measurements can generate non-classical correlations in the last two qubits of the ladder-like cluster state. Our study also reveals the flow of non-classical correlations within the cluster state as the measurements are progressively done, throwing light on the dynamics of such correlations during the execution of a quantum algorithm within the MBQC paradigm. We see that with random measurements on $n-3$ qubits in an $n$ qubit ladder, the non-classical correlations between the last two qubits grow stronger on an average when $n$ and the number of rungs of the ladder increases. We also find that with more rungs in the ladder, the last two qubits start to share bipartite entanglement in addition to other more generic non-classical correlations. With only a single non-Clifford measurement, we find that the flow of correlations through the ladder is quite non-intuitive, highlighting the different and unexpected ways in which delocalized quantum information in the ladder can be modulated and moved around through measurements on various qubits on the rungs of the ladder. 

For the small clusters with 4 and 6 qubits respectively we were able to understand the behavior of the system under local qubit measurements by computing how the stabilizer generators transform. For both cases, we were able to see that for certain choices of measurement angles, the state of the remaining cluster after the measurements remains a stabiliser state and there will not be any non-classical correlations developed between the two qubits in the last rung. The choices of these measurements are very specific and in almost all cases, the resultant cluster is not a stabilizer state and non-classical correlations are built-up across the last two qubits. This result builds on those in~\cite{SubsystemDiscord} wherein a generic connection between multipartite entanglement in structured states like cluster states and subsystem discord was established. 

Our results suggests a dynamic interplay between the structure of the cluster state and the influence of local measurements. Such insights not only contribute to the understanding of the behavior of non-classical correlations in cluster states but also hold significance for the optimization and design of quantum computing architectures, particularly those employing cluster states for MBQC applications. If the last two qubits are viewed as an open quantum system, our results show how starting from an uncorrelated state, the two can get correlated through operations - more specifically, measurements - being performed on other parts of the ladder state which form the environment of the two qubits.

\section*{Acknowledgements}
A.~S.~was supported by QuEST grant No Q-113 of the Department of Science and Technology, Government of India. The authors acknowledge the centre for high performance computing of IISER TVM for the use of the HPC cluster, {\em Padmanabha}. 

\bibliography{InformationFlow}

\end{document}